\documentclass[aps,pre,twocolumn,showpacs,superscriptaddress]{revtex4-1}
\usepackage{bm}%
\usepackage{float}
\expandafter\ifx\csname package@font\endcsname\relax\else
 \expandafter\expandafter
 \expandafter\usepackage
 \expandafter\expandafter
 \expandafter{\csname package@font\endcsname}%
\fi

\usepackage{array}
\usepackage{amsmath,amssymb}
\usepackage{MnSymbol}
 \usepackage{tikz-feynman} 
\usepackage{dcolumn}
\usepackage{graphics}
\usepackage{graphicx}
\usepackage{mathrsfs}
\usepackage{xcolor}
\usepackage{mathtools}
\usepackage{epstopdf}
\usepackage{hyperref}
\usepackage{verbatim} 

\hypersetup{%
   pdfpagemode=None, 
   pdfstartpage=1,
   pdfmenubar=true,
   pdftoolbar=true,
   colorlinks = true,
   linkcolor=blue,
   citecolor=blue,
   urlcolor=blue,
   bookmarksopen=false
 }
 \usepackage{amsmath}
\usepackage{amsfonts}
\usepackage{mathtools}
\usepackage{graphicx}
\usepackage{fixme}
\usepackage{color}
\usepackage{soul}
\usepackage[section]{placeins}
\usepackage{enumitem}
\usepackage[normalem]{ulem}

\begin{document} 
\title{Chaos and Regularity in an Anisotropic Soft Squircle Billiard}
\author{A. González Andrade}
\affiliation{Departamento de F\'isica, Universidad Aut\'onoma Metropolitana-Iztapalapa, Av. San Rafael Atlixco 186, C.P. 09310 CDMX, M\'{e}xico}
\author{H. N. Núñez-Yépez}
\affiliation{Departamento de F\'isica, Universidad Aut\'onoma Metropolitana-Iztapalapa, Av. San Rafael Atlixco 186, C.P. 09310 CDMX, M\'{e}xico}
\author{M. A. Bastarrachea-Magnani}
\email{bastarrachea@xanum.uam.mx}
\affiliation{Departamento de F\'isica, Universidad Aut\'onoma Metropolitana-Iztapalapa, Av. San Rafael Atlixco 186, C.P. 09310 CDMX, M\'{e}xico}

\begin{abstract}
A hard-wall billiard is a mathematical model describing the confinement of a free particle that collides specularly and instantaneously with boundaries and discontinuities. Soft billiards are a generalization that includes a smooth boundary whose dynamics are governed by Hamiltonian equations and overcome overly simplistic representations. We study the dynamical features of an anisotropic soft-wall squircle billiard. This curve is a geometric figure that seamlessly blends the angularity of a square with the smooth curves of a circle. We characterize the billiard's emerging trajectories, exhibiting the onset of chaos and its alternation with regularity in the parameter space. We characterize the transition to chaos and the stabilization of the dynamics by revealing the nonlinearity of the parameters (squareness, ellipticity, and hardness) via the computation of Poincar\'e surfaces of section and the Lyapunov exponent across the parameter space. We expect our work to introduce a valuable tool to increase understanding of the onset of chaos in soft billiards.
\end{abstract} 

\maketitle

\section{Introduction}

A dynamical billiard is a paradigmatic model that exemplifies a conservative dynamical system~\cite{Datseris2022}. These models describe the physical behavior of one or more point particles moving freely within a bounded domain, e.g., an infinite step potential well, where the particles interact by colliding with the domain’s boundaries or walls~\cite{Chang1993}. The collision is elastic, and because it instantaneously changes the sign of the particle's normal component, it is called specular reflection~\cite{Tabachnikov2005}. The dynamical properties of such systems are determined by the shape of the boundaries, which can range from entirely regular (integrable) to fully chaotic~\cite{Chernov2006, Bunimovich2019}, making them an appropriate model for studying chaotic dynamics. We can also refer to these dynamical systems as hard-wall billiards. Billiards are paradigmatic models for the study of dynamical chaos~\cite{Ott2002}, the emergent phenomena characterized by the sensitivity to initial conditions in phase space. This is the case of notorious examples such as the Sinai billiard and the Bunimovich stadium billiard, well known for their chaotic dynamics~\cite{Chernov2006}. 

In this context, {\it soft billiards} emerge as a generalization of the standard hard-wall billiards whose boundaries are defined by a non-step potential well. They have been employed before for studying many features such as the robustness of trajectories in the hard-wall limit~\cite{Rapoport2007}, the appearance of regular islands in phase space~\cite{Turaev1998,RomKedar1999}, hyperbolicity~\cite{Balint2006}, the inducement of stickiness~\cite{Custodio2010}, the impact of corners~\cite{Turaev2003}, in the dynamics of Lorentz gas models~\cite{Baldwin1988,Knauf1989}, the three-degrees-of-freedom scattering problem~\cite{Drotos2014}, soft impacts~\cite{Kloc2014} and even quantum scarring~\cite{Luukko2016}. Previous studies have dealt with some geometries, including the 1D billiards~\cite{Oliveira2008}, elliptic billiard~\cite{Kroetz2016}, and a triangular one where stickiness has been revealed~\cite{Oliveira2012}. Moreover, they have been realized experimentally in a tilted Bunimovich stadium billiard employing a cold atoms setup~\cite{Kaplan2001}, quantum dots~\cite{Rotter2003,Weingartner2005}, as a soft-walled microwave resonator~\cite{Kim2005}, and as open microwave cavities~\cite{Kim2005}. In addition, there are cutting-edge experimental realizations of billiard systems such as flat, metallic microwave resonators, where the transition to chaos has been investigated~\cite{Zhang2019}, or exciton-polariton billiards~\cite{Gao2015}. It has been demonstrated that incorporating soft walls into a billiard system results in a mixed phase space~\cite{Sachrajda1998,Kim2005}, either introducing chaos in regular setups or stabilizing the dynamics and creating stability islands~\cite{Kaplan2001,Kroetz2016}. However, many questions remain open, as the effects of introducing and tuning wall smoothness on the dynamics have not been extensively studied. Given the diversity of experimental realizations of soft billiards and their wide-ranging applications in technology and development, they stand as suitable models for in-depth studies of the onset of chaos, a guiding principle for the present work. 

In this contribution, we investigate the dynamical characteristics of a soft billiard with a tunable geometry: an anisotropic squircle. This particular geometry comes as a generalization of the ellipses called superellipses $(x/a)^{n} + (y/b)^{n} = 1$. For $n\geq 2$, $n$ being a positive integer, we get a family of curves called Lam\'e curves~\cite{Gridgeman1970}, and when $a=b$ we have the supercircles. We focus on a specific parametrization of supercircles, the {\it squircle}~\cite{Guasti1992}, which enables a continuous transition from a circle to a square, with the supercircle between them. Subsequently, we add an anisotropy that transforms the squircle into a superelliptic shape. The squareness and anisotropy parameters provide access to a broader parameter space, enabling the exploration of the onset of chaos and regularity introduced by smoothness and its interaction with geometry. Notably, we find a nonmonotonical presence of regular oscillations of the amount of chaos as a function of the parameters, a phenomenon known elsewhere as {\it breathing of chaos}~\cite{Richter1990,Dullin1996}. Few have been studied about these curves in the hard-wall limit, including the problem of a particle confined in a supercircular box~\cite{Bera2008}, and planar optical microresonators~\cite{Burke2019,Boriskina2004}, but also serves as a good approximation to realistic setups without sharp corners, which are experimentally challenging to reach~\cite{Huang2002}. Also, the resulting potential has strong similarities to Caldera-like potentials~\cite{Katsanikas2020}, used to model chemical reactions.

The article is organized as follows. In Sec.~\ref{sec:2}, we describe the Hamiltonian of a soft billiard with a general anisotropic squircle shape, its dynamical equations, and how we solve them. Then, in Sec.~\ref{sec:3}, we analyze the different orbits emerging as a function of hardness and squareness in the squircle. Next, in Sec.~\ref{sec:4}, we discuss how to determine and identify collisions in soft billiards and phase space portraits. In Sec.~\ref{sec:5}, we quantify chaos by comparing the average and maximal Lyapunov exponent over the parameter space of the Hamiltonian. Finally, in Sec.~\ref{sec:6}, we offer our perspectives and conclusions.

\section{Hamiltonian}
\label{sec:2}

\begin{figure}
    \centering
    \includegraphics[width=0.45 \textwidth]{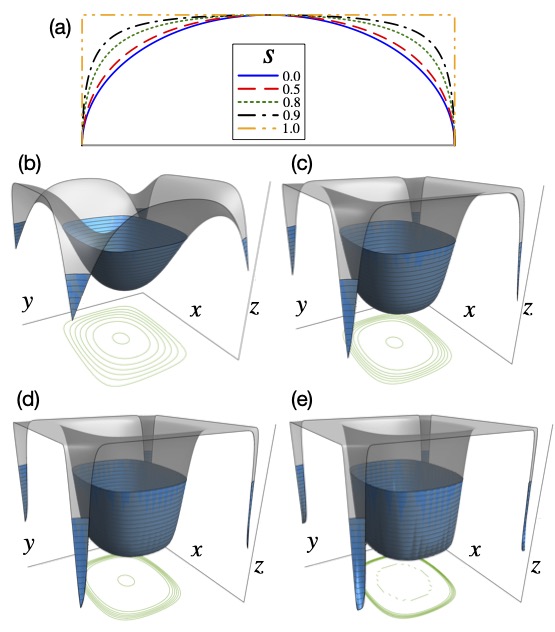}
    \caption{(a) The squircle curve is plotted for various squareness values $s$ for a fixed value $\tilde{\beta} = 0.7$. Below is depicted the surface corresponding to the equation \eqref{ec:Hamiltoniano_Billar_Squircle}, for different hardness values given by (b) $ h=1$, (c) $h=3$, (d) $h=5$, (e) $h=10$. With $s=0.8$ and $\tilde{\beta}=0.8$ fixed. The energy is chosen to be $E=0$, represented as the blue surface. The green lines at the bottom represent the equipotential curves where collisions can occur.}
    \label{fig:1}
\end{figure}

We consider a billiard whose border, in the hard-wall limit, is defined by a squircle curve in the plane. The \textit{squareness parameter} $s$ is associated with the sharpness of a square, and it produces edges and fillets of any arbitrary degree of roundness~\cite{Barr1981}. This geometric parameter allows us to transition from a circumference to a square and vice versa. For the $s=0$ ($s=1$) limit, we get an ellipse (rectangle) as shown in Fig.~\ref{fig:1} (a). Moreover, we introduce the anisotropic case whose equation is~\cite{Guasti1992}
\begin{gather}
    \frac{x^2}{\alpha^2} + \frac{y^2}{\beta^2} - \frac{s^2 x^2 y^2}{\alpha^2 \beta^2} = 1 \, ,
\end{gather}
where $\alpha$ and $\beta$ are the major and minor semiaxes, respectively. Notice that, for soft billiards, the curve does not define the billiard table, just the energy reference with respect to the hard-wall case.

It has been demonstrated that a substantial class of smooth potentials can model a soft-wall billiard~\cite{Turaev1998}. We define a softened billiard-like potential that models a squircle as well as
\begin{gather}\label{eq:Soft_Potential}
    V (x,y;h) = \text{erf}\left[h\left(\dfrac{x^2}{\alpha ^2} + \frac{y^2}{\beta^2} - \frac{s^2 x^2 y^2}{\alpha ^2 \beta^2}\right)\right]
\end{gather}
where $\text{erf}(z) = (2\pi)^{-1/2} \int_0 ^z e^{-t^2} dt$ is the error function and $h\in(0,\infty)$ is the \textit{hardness parameter}, which softens the profile in the walls of the well. Generally, for the limit $h\rightarrow \infty$, the Hamiltonian flow converges to the billiard flow, which has been proved before~\cite{Turaev1998}. This function is selected based on the Gaussian profile observed in potential barriers in the laboratory~\cite{Gao2015}. Although there are alternative methods for softening the walls of the billiards, such as polynomial functions~\cite{Ortiz2000}, the error function remains the preferred choice due to its well-established properties. Also, Fermi potentials are the preferred choice in the context of Lorenz gases~\cite{Klages2019,Gil-Gallegos2019,Toivonen2025,Toivonen2025arXiv}. The potential in Eq.~\eqref{eq:Soft_Potential} is plotted in Fig.~\ref{fig:1} (b)-(e). The soft billiard corresponds to the two-dimensional flat surface onto which the equipotential surfaces are projected. Additionally, the region of permitted motion in the configuration space is depicted in blue. We have identified this region as the Hill’s region for the soft billiard~\cite{Rapoport2008}. The set of points where a particle attains a critical energy corresponds to the zero velocity curve, projected as the largest equipotential in the billiard case.

In the context of a horizontal squircle billiard, we employ a dimensionless formulation of the equations, ensuring that the potential explicitly depends on the variable $\beta$. This variable is designated as the free semi-axis, drawing parallels to the terminology employed for ellipses. By substituting $x=\alpha\tilde{x}$,  $y=\alpha\tilde{y}$, and $\beta=\alpha\tilde{\beta}$, the squircle billiard's Hamiltonian reads
\begin{gather}\label{ec:Hamiltoniano_Billar_Squircle}
    H(\tilde{x},\tilde{y};h,s,\tilde{\beta}) = \frac{1}{2}(p_{\tilde{x}}^{2} + p_{\tilde{y}}^{2}) + \text{erf}\left[h\, (Q -1)\right] \, ,
\end{gather}
where 
$Q = \tilde{x}^{2} + \tilde{y}^{2}/\tilde{\beta}^{2} -s^2 \tilde{x}^2 \tilde{y}^2 / \tilde{\beta}^2$. The Hamilton equations for the dynamics of a unitary mass particle in the soft squircle billiard are
    \begin{gather}\label{eq:Hamilton_Equations_Squircle}
        \frac{d\tilde{x}}{dt} = p_{\tilde{x}},\,\,\,\, 
        \frac{d\tilde{y}}{dt} = p_{\tilde{y}},
    \end{gather}
    \begin{gather}\label{eq:Hamilton_Equations_Squircle2}
        \dfrac{dp_{\tilde{x}}}{dt} = -\dfrac{4h\tilde{x}}{\sqrt{\pi}} \left(1-\frac{s^2 \tilde{y}^2}{\tilde{\beta}^2} \right) \exp \left[ -h^2 (Q-1)^2 \right] \, ,
    \end{gather}
    \begin{gather}\label{eq:Hamilton_Equations_Squircle3}
        \dfrac{dp_{\tilde{y}}}{dt}  = -\dfrac{4h\tilde{y}}{\tilde{\beta}^2\sqrt{\pi}} \left(1-s^2 \tilde{x}^2 \right)  \exp \left[ -h^2 (Q-1)^2 \right] \, .
    \end{gather}
This set of differential equations presents stiffness as $h$ increases. We use an adaptive high-order and stage Runge-Kutta method of six stages and fifth algebraic order to deal with this issue~\cite{Rackauckas2017,Tsitouras2019}. The initial step accounts for the proximity of the particle to the boundaries. If the particle is near the boundaries, the initial step is of the order $10^{-5}$. If not, it is of the order $10^{-3}$. Using the former approach, the adaptive method can deal with stiff differential equations and numerical overflow is avoided, albeit at the expense of increased computing times.

\section{Geometrical limits and dynamics}
\label{sec:3}

\begin{figure}
    \centering
    \includegraphics[width=0.8\columnwidth]{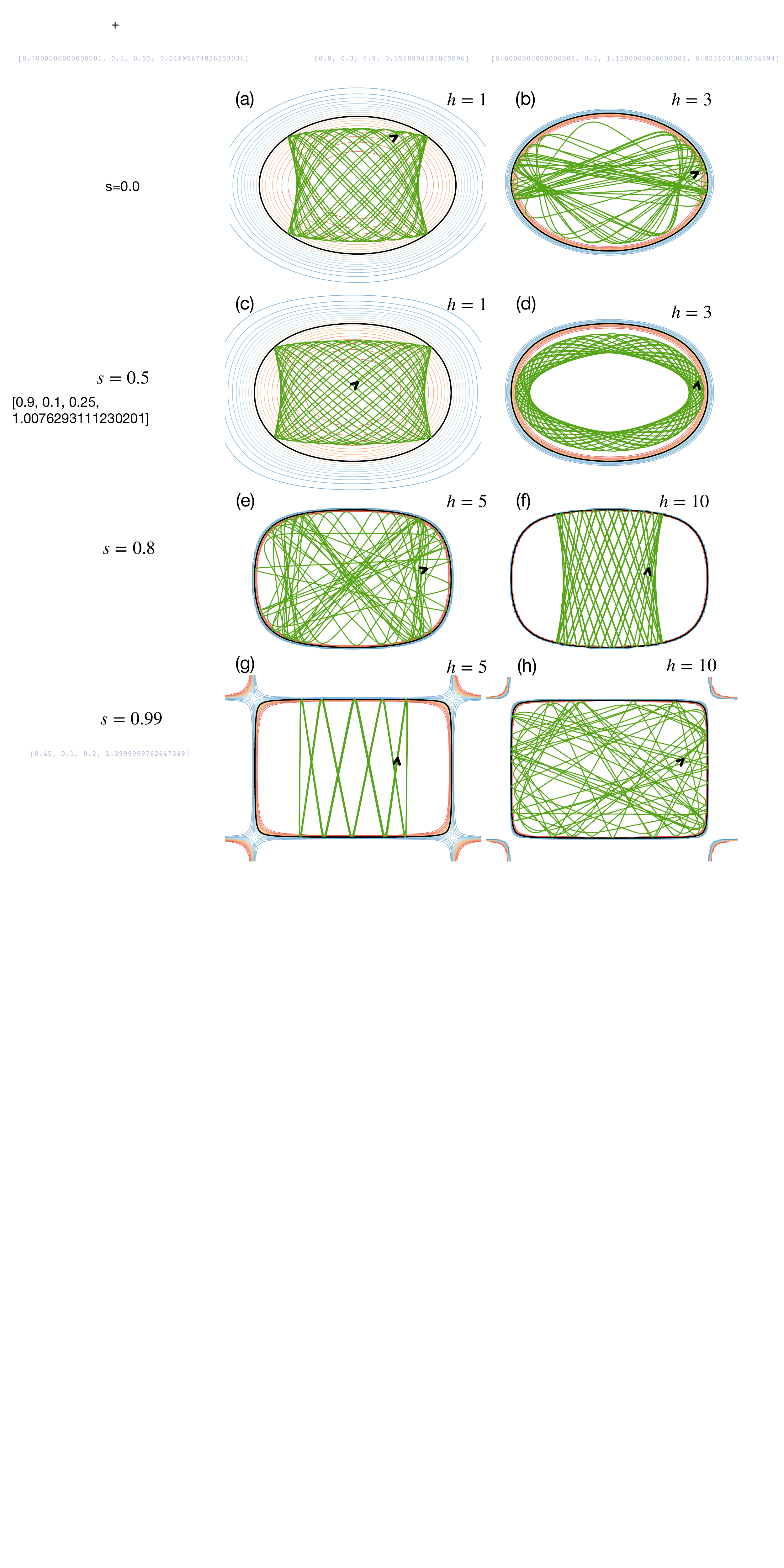}
    \caption{Sample trajectories (green) in the squircle soft billiard are shown. Parameters $\tilde{\beta}=0.7$ and $E=0$ (black curve associated) are fixed. Equipotential curves are plotted (in a gradient from orange to blue) to exhibit the behavior of the soft potential as $h$ and $s$ parameters vary. The values of squareness are fixed horizontally as (a), (b) $s=0.0$, (c), (d) $s=0.5$, (e), (f) $s=0.8$ and (g), (h) $s=0.99$.}
    \label{fig:2}
\end{figure}

In Fig.~\ref{fig:2}, we show representative trajectories for a particle in the anisotropic squircle billiard obtained by solving Eqs.~\eqref{eq:Hamilton_Equations_Squircle} to~\eqref{eq:Hamilton_Equations_Squircle3}. The black curve represents the zero-energy equipotential surface, corresponding to the hard-wall boundary condition. In each plot, we select an arbitrary zero-energy initial condition, indicated by an arrow. 

In the first row of Fig.~\ref{fig:2}, we have the elliptic case ($s=0$), whose dynamics are regular in the hard-wall limit~\cite{Tabachnikov2005}. As expected, variations only in the hardness parameter change the dynamics, inducing chaos~\cite{Kroetz2016}. This behavior persists as we vary the value of $s$, suggesting that the susceptibility of the dynamics to the hardness parameter extends beyond typical elliptical billiards. Subsequently, as we choose larger values for $s$, we observe that, analogously to the hardness parameter, the dynamics exhibit sensitivity to variations in squareness. It can be qualitatively seen that a gradual increase of the hardness parameter either increases or decreases the regularity in the configuration space trajectories. This dynamical behavior is related to a nonlinear increase of chaotic domains in phase space with respect to the  parameters~\cite{Custodio2010,Custodio2012}. As we will observe later by studying the Lyapunov exponent, the interplay between the round corners that induce chaos~\cite{Custodio2012}, and the stabilization effects due to softness will create average stripes of chaos in the parameter space, associated with the phenomenon known as breathing of chaos.

In an elliptical billiard~\cite{Kroetz2016}, two distinct movement regimes can be identified: rotational and librational. The first type of orbit corresponds to trajectories that form an envelope around the foci (in the elliptical case). In contrast, the second type involves crossing the horizontal axis through the interfocal line~\cite{Riestra2024}. This type of movement can be identified in the regular trajectories for the squircle billiard without losing generality.

\section{Collisions}
\label{sec:4}

\begin{figure}
    \centering
    \includegraphics[width=0.6\columnwidth]{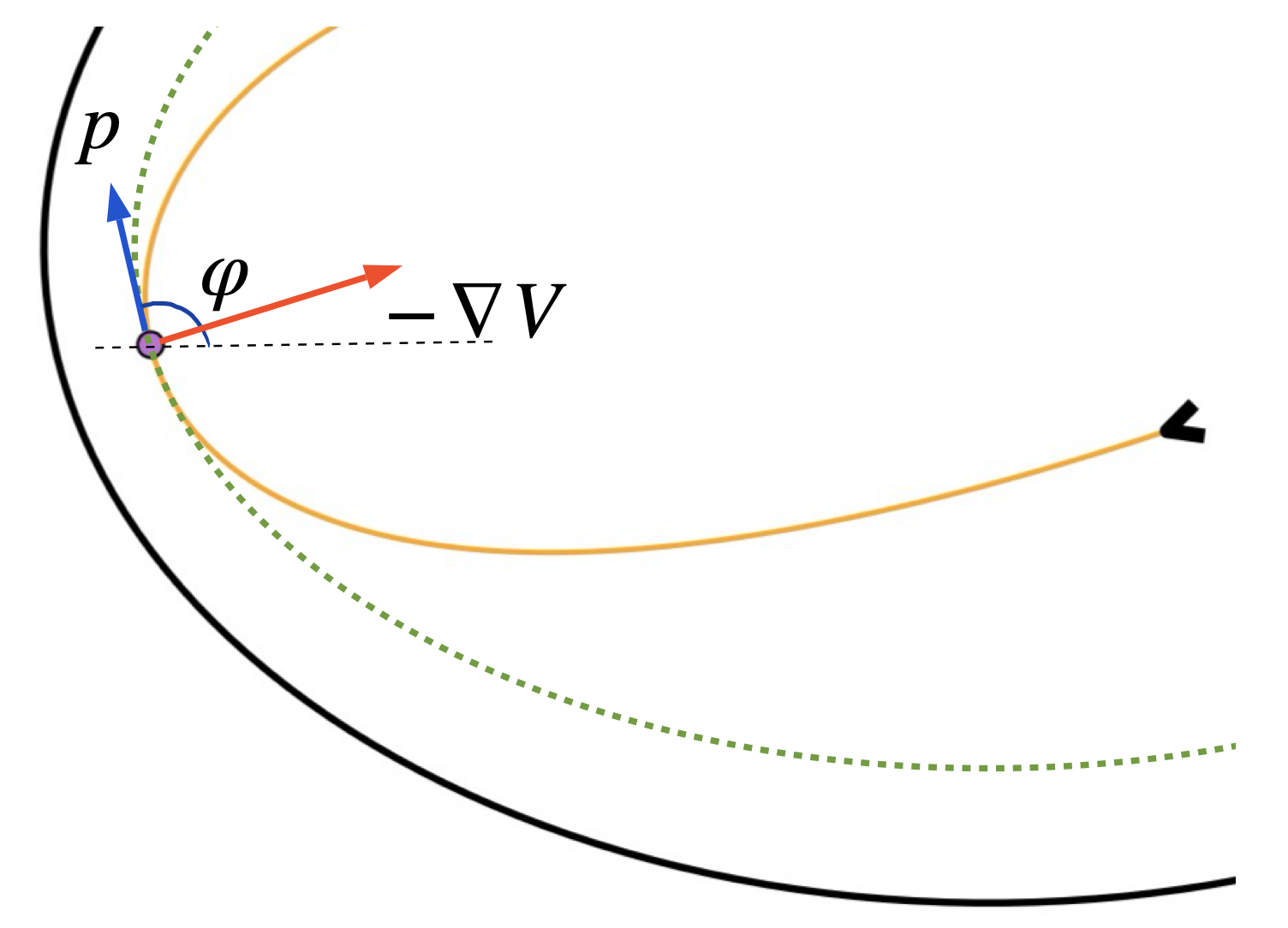}
    \caption{The instant of collision is defined when the moment becomes entirely parallel to the equipotential curve (green dashed line). This moment forms an angle $\varphi$ with the horizontal. The particle’s trajectory is depicted in orange, the equipotential curve for $E=0$ is in black, and the direction of the potential force is in red.}
    \label{fig:colision_definicion}
\end{figure}

To perform a qualitative study of the dynamics through Poincar\'{e} surfaces of sections (PSOS), we employ what is known as the collisions space~\cite{Chernov2006}. However, for a soft-wall billiard, redefining the concept of collision is necessary. 

The dynamics of a hard-wall billiard are defined by the way the particles scatter from the boundary. Let us denote $v$ the velocity of the particle before the collision, $v'$ the velocity after it, $\mathbf{n}$ the normal vector, and $\mathcal{T}$ the tangent vector, both at the point of impact. So we assume elastic ($v \mathcal{T} = v' \mathcal{T}$) and specular ($v \mathbf{n} = -v' \mathbf{n}$) reflections at the point of collision. In this case, the dynamics are governed by mappings that retain all the important features and intrinsic complexities of a continuous evolution~\cite{Smilansky1992}. However, for smooth boundaries where such discontinuities in the dynamics do not exist, it is necessary to numerically find the state of the particle at the instant of the collision, which means the point where the particle reaches an equipotential curve and consequently changes the direction of its momentum. The soft billiard potential always seeks to return the particle to the center. Thus, it is established that the collision occurs when the particle's momentum is tangent to the equipotential curve, i.e., perpendicular to the direction of the force (see Fig.~\ref{fig:colision_definicion}). Thus, for the smooth \textit{squircle} billiard, the collision condition is~\cite{Kroetz2016}
\begin{gather}\label{ec:Condicion_Colision_Squircle}
     \mathbf{p} \cdot \mathbf{F} = \mathbf{p} \cdot (- \nabla V) = \tilde{x} p_{\tilde{x}} + \dfrac{\tilde{y} p_{\tilde{y}}}{\tilde{\beta}^2} - \dfrac{s^2 \tilde{x} \tilde{y}}{\tilde{\beta}^2} (\tilde{x}p_{\tilde{y}}+\tilde{y}p_{\tilde{x}}) = 0.
\end{gather}

In the hard-wall limit, the pair of coordinates that defines the collision space is called the Poincaré-Birkhoff coordinates ($\xi, v_t$)~\cite{Lozej2021}, where $\xi$ is the arc length of the boundary, and $v_t$ is the tangential component of the velocity at the collision instant. Given that in a soft-wall billiard collisions  occur with an arbitrary number of equipotential curves, we define the $(\theta, p_t)$ as the Poincaré-Birkhoff coordinates for the soft-wall billiard~\cite{Kroetz2016}, where $\theta$ is the polar angle and $p_t$ is the momentum of the particle at the collision which in fact is tangential to the equipotential curve. 

\begin{figure}[t]
    \centering
    \includegraphics[width=0.99\columnwidth]{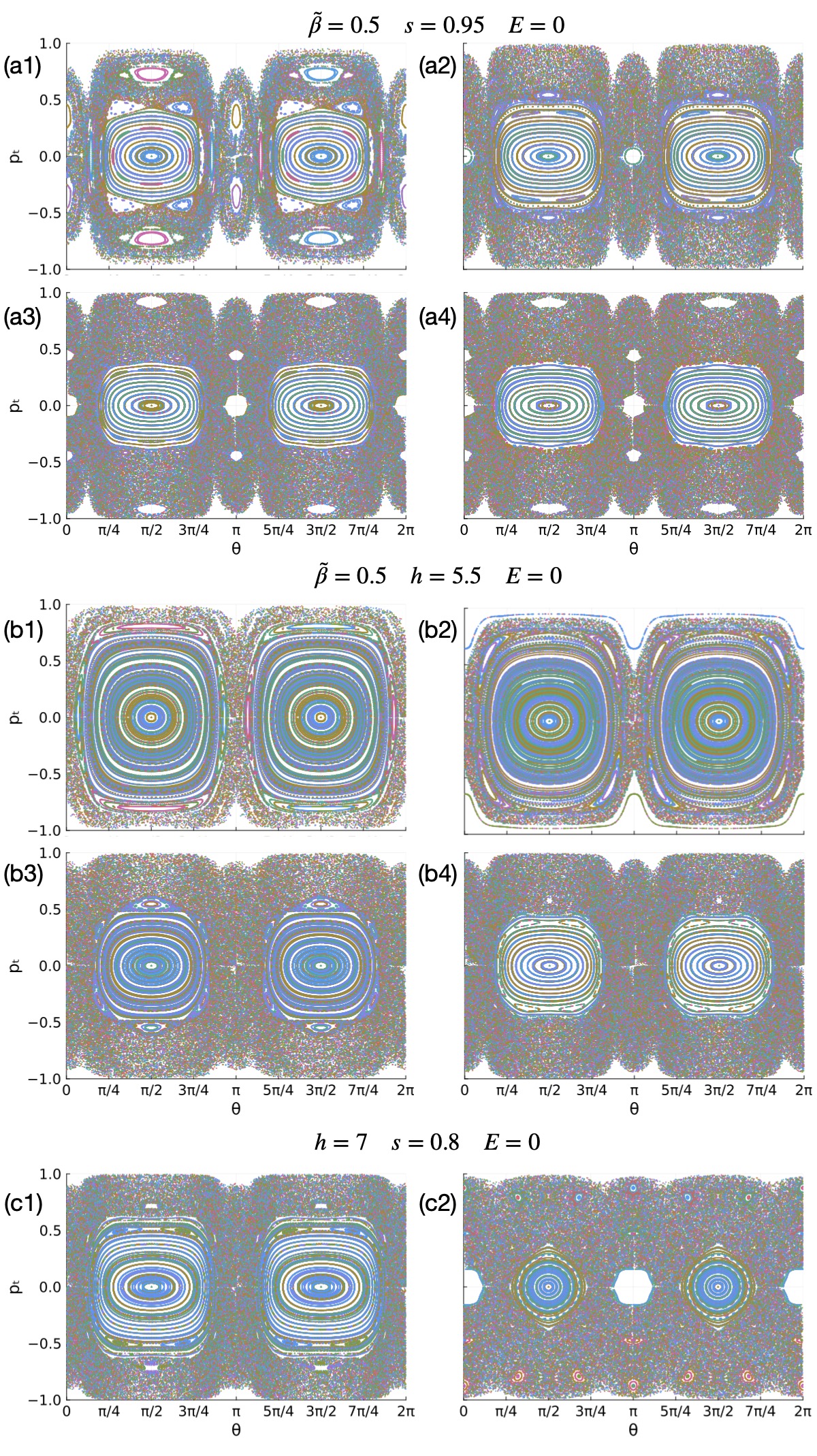}
    \caption{PSOS for the soft squircle billiard from a set of random initial conditions. (Top) for fixed $s=0.95$, $\tilde{\beta}=0.5$, and hardness values are chosen as: (a1) $h=2$, (a2) $h=4$, (a3) $s=6$, (a4) $h=8$. (Middle) for fixed $\tilde{\beta}=0.5$, $h=5.5$, and squareness values (b1) $s=0.3$, (b2) $s=0.5$, (b3) $s=0.8$, (b4) $s=0.9$. (Bottom) for fixed $s=0.8$ y $h=7$, and values of the free semiaxis (c1) $\tilde{\beta}=0.4$ and (c2) $\tilde{\beta}=0.8$.}
    \label{fig:4}
\end{figure}

\section{Chaos and Regularity}
\label{sec:5}

In this section, we study the dynamics of the soft-wall billiard as a function of its parameters both qualitatively and quantitatively.

\subsection{Poincar\'e Surfaces of Section}

We construct the PSOS by taking a representative sample of initial conditions, i.e., sweeping $\tilde{x}$ and $\tilde{y}$ coordinates and assigning a random value to the momenta. We notice that richer dynamics appear for initial conditions closer to the borders $\tilde{x}\rightarrow \pm 1$ and $\tilde{y} \rightarrow \pm \tilde{\beta}$ (zero velocity curve). 

\begin{figure*}
    \centering
    \includegraphics[width=1.0\textwidth]{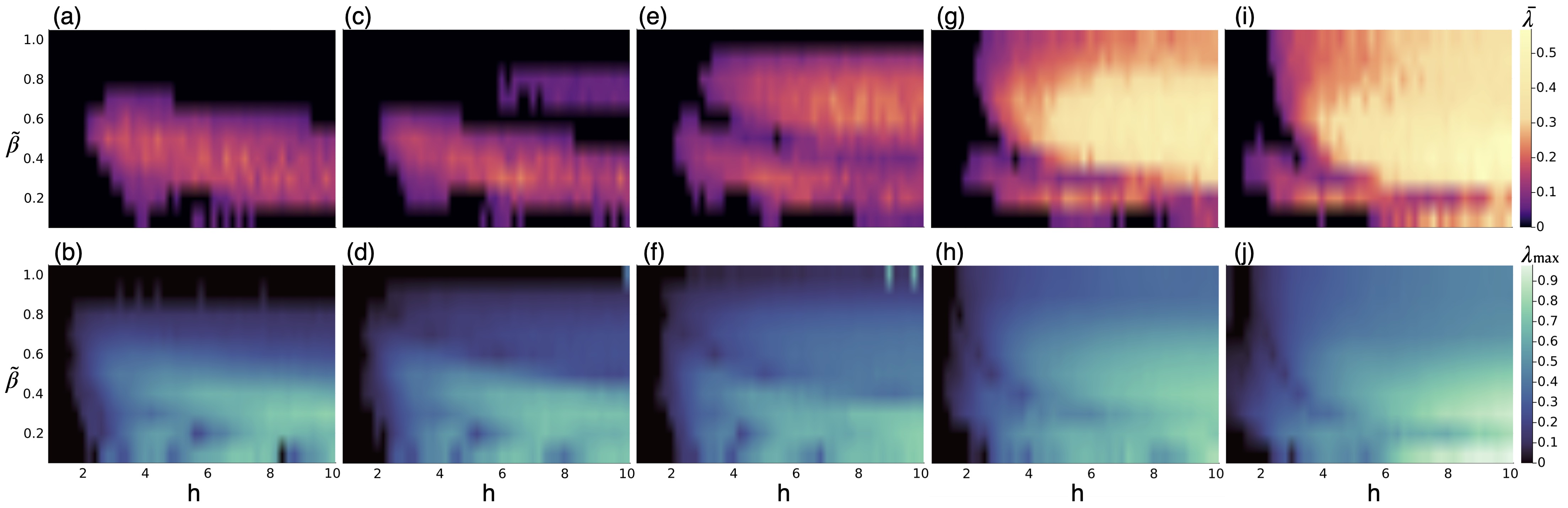}
    \caption{MLE for a random set of initial conditions as a function of $h$ and $\tilde{\beta}$ for $E=0$. (Above) $\bar{\lambda}$ and (below) $\lambda_{\text{max}}$ with $s$ (from left to right): (a)-(b) $ s=0.0$, (c)-(d) $s=0.3$, (e)-(f) $s=0.5$, (g)-(h) $s=0.8$, (i)-(j) $s=0.9$ with $E=0$ fixed.}
    \label{fig:5}
\end{figure*}

First, in Figs.~\ref{fig:4} (a1), we notice the presence of regular regions occupying larger phase space domains, usually near the multiples of $\pi/2$ of the $\theta$ angle. This is associated with trajectories far from the corners, which are relevant for a squircle billiard, unlike the elliptical case. In Fig.~\ref{fig:4} (a2), many stability regions disappeared, which shows that increasing hardness makes chaos predominance notable. Nevertheless, as we can observe in Fig.~\ref{fig:4} (a3), regular regions emerge again but in different places of phase space. Finally, in Fig.~\ref{fig:4} (a4), we observe that as we increase the hardness, the dynamics do not present significant changes in the size and spatial distribution of regularity regions. This suggests the existence of a certain $h$ for which the dynamics tend toward stability.

\begin{figure*}
    \centering
    \includegraphics[width=1\textwidth]{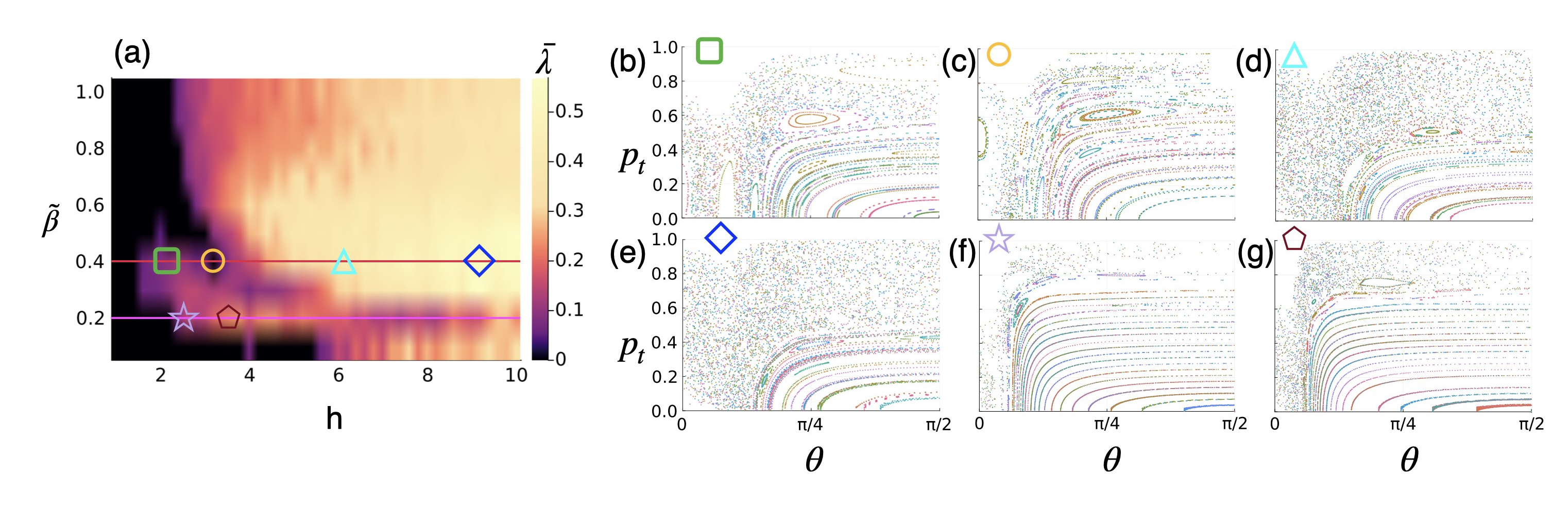}
    \caption{(a) Average MLE for a random set of initial conditions as a function of $h$ and $\tilde{\beta}$ and $s=0.9$ (b) to (g) PSOS for the same initial conditions. With $\tilde{\beta}=0.4$ [(b) to (e)], and $\tilde{\beta}=$0.2 [(f) to (g)] and the values of $h$ indicated by the figures. This enables us to observe the transformation of the chaotic sea as the hardness parameter undergoes its changes.}
    \label{fig:5a} 
\end{figure*}
PSOS for different values of the squareness parameter of the squircle are shown in Figs.~\ref{fig:4} (b1)-(b4). In this case, it is observed that the gradual increase in the squareness promotes a corresponding gradual onset of chaos. In Figs.~\ref{fig:4} (b1) and (b2), the dynamics exhibit similar behavior, with the phase space area associated with non-regular trajectories being comparable in both cases. Conversely, for a square closer to the rectangular case, the dynamics are chaotic, except in the vicinity of $p_t = 0$ for $\theta = \pi/2$, $3\pi/2$. This figure demonstrates that the dynamics of the square billiard are also sensitive to the parameter $s$, which deforms the chaotic region and encompasses the regularity.

A fundamental aspect of the study of hard-wall billiards is that the geometry of the boundary governs the dynamics. Consequently, it is expected that altering the shape of a soft-walled billiard will reconfigure the structure of phase-space, enlarging some regular islands and expanding the chaotic sea elsewhere. This is illustrated in Figs.~\ref{fig:4} (c1)-(c2), where both stable and unstable regions are observed for the case $\tilde{\beta} = 0.4$ (closer to the ellipse). In contrast, for the case $\tilde{\beta} = 0.8$ (closer to the rectangle), the size of the regular regions changes, and new regions also emerge at $\theta = \pi, 2\pi$. Therefore, the dynamics of soft-walled billiards are also sensitive to the geometry of the boundary beyond the hardness it presents. Altogether, hardness ($h$) and geometry ($\tilde{\beta}$, $s$) allow for the modulation of chaos in a system, enabling chaos-regularity-chaos transitions. Consequently, a notable advantage of soft-walled billiards, especially those with squircle boundaries, is that with a single numerical tool, various aspects of the dynamics can be studied by simply adjusting the appropriate parameters.

\subsection{Maximum Lyapunov Exponent}

We compute the Maximum Lyapunov Exponent (MLE)~\cite{dellago1996,Ott2002} as a function of the parameter space. Random initial conditions are swept to ensure they form a representative phase space sample. This is accomplished by generating initial values for each position coordinate, first $\tilde{x}$ and then $\tilde{y}$. Next, each element is randomly assigned the remaining values that determine the initial state of the particle ($\tilde{x}_0$, $\tilde{y}_0$, $p_{\tilde{x},0}$, $p_{\tilde{y},0}$), always ensuring that they maintain the same energy value ($E = 0$), with a final mesh of 12x9 random initial conditions for the one quadrant of billiard. The dynamics  evolve for all the initial conditions until the MLE converges (usually around $4 \times 10^4$ time steps). We use the same adaptive method described before, combined with the “DynamicalSystems.jl” library of the {\it Julia} language, to compute the MLE. 

In one case, the average exponent among these initial conditions is calculated, denoted as $\bar{\lambda}$. On the other hand, the maximum value that the Lyapunov exponent can reach among them is calculated, labeled as $\lambda_{\text{max}}$. From this parameter space, it becomes easy to locate initial conditions that generate highly chaotic trajectories (for $\lambda _{max}$) or, at least, identify regions where they are more likely to be found (for $\bar{\lambda}$).

In the first row of Fig.~\ref{fig:5}, we show the parameter space of $\tilde{\beta}$ and the hardness as the squareness of the boundary varies for fixed eccentricity. Notably, chaos becomes predominant on average as squareness increases (from $s=0$ to $0.9$).  For $s=0$, in Fig.~\ref{fig:5} (a), only a streak of chaos appears across the parameter space introduced by the billiard's softness. There, we observe stripes-like structures with non-negligible variations of the order of 0.1 in the Lyapunov exponent as a function of $h$. Billiards have been studied considering a nonlinear parameter, resulting in stripe-like structures like these when plotting initial conditions against the parameter. This intermittent behavior between chaos and regularity, which can also be glimpsed in the Poincaré sections of Fig.~\ref{fig:5} (b), seems characteristic of integrable Hamiltonian systems where weak chaos is introduced~\cite{Custodio2012}. Then, as $s$ increases, a second streak of chaos begins to form on the map. We attribute this behavior to the presence of the corners. Indeed, increasing $s$ makes this region larger until it converges with the original one, as seen from Figs.~\ref{fig:5} (c) to (i), forming a large stripe of regularity separating them. This regular band persists with the increase of squareness. We observe that the stripe-like structures appear in both chaotic regions, but starts to vanish for large values of $h$ and $s$, when the ergodicity dominates, as discussed below. Although it is not shown in Fig.~\ref{fig:5} (g)-(j),  increasing hardness progressively makes elliptic geometries turn more regular, as expected.

In the row below of Fig.~\ref{fig:5}, the maximum value of the MLE obtained for each combination of the same parameters is shown, unveiling the same nonlinear behavior, except that the intermittence between regularity and chaos is not observed in the maximum value of the MLE. Unlike the average MLE $\bar{\lambda}$, this maximum value depends on the timescales of the system dynamics, so it signals the gradual, increasing growth of chaos as one modifies the parameters~\cite{Nakerst2023}. Upon analyzing the parameter space, it becomes apparent that increasing the squareness parameter makes the overall behavior more unstable. This is shown by the fact that for values of $s = 0$, $0.3$, $0.5$, $0.8$, the maximum value of the MLE is around $\lambda_{\text{max}} \approx$ $0.9$. For the maximum possible value of the Lyapunov exponent, the distribution in parameter space remains nearly unchanged until the value $s=0.9$, when it attains its maximum value in many points, reaching ergodic behavior.

The results for both the average and maximal MLE are consistent with the observations previously obtained from the Poincaré sections. That is, as we keep a fixed value of the free semi-axis, varying the hardness leads to regularity-chaos-regularity transitions. Consequently, it becomes apparent that the dynamics of a soft squircle billiard exhibit a heightened sensitivity to the squareness and hardness of the boundary. 

In Fig.~\ref{fig:5a}, we follow the behavior of the PSOS across the parameter space (fixing $\bar{\beta}=0.4$) to observe in detail the effect of the stripes and the regular band in phase space. From Fig.~\ref{fig:5a} (b) to (e), we directly observe the changes in the size of the chaotic sea as stability islands emerge and recede until one attains the ergodic regime, where they vanish. From low to high values of $h$ one crosses from chaotic regions across the band of regularity. The transition is manifested in a shift in the dominance of circular over square behavior on the billiard, where the phase space occupied by rotational orbits decreases.

Then, in Fig.~\ref{fig:5a} (f) and (g), we show how the stripes as a function of softness are expressed as the vanishing and emergence of stability islands, thus changing the size of chaotic regions of phase space. Similar behavior to the stripes was observed in the soft elliptic billiard~\cite{Kroetz2016}, which was solely associated with the hardness parameter, as well as in other systems such as wedge ~\cite{Richter1990}, and Benettin and Strelcyn billiards~\cite{Dullin1996}, and it is generally known as breathing of chaos~\cite{Richter1990}. For the soft elliptical billiard it  was attributed to rearranging the available phase space for the two types of orbits: libration and rotation~\cite{Kroetz2016}. Also, these stripes might represent a predominance of particular sets of initial conditions associated with chaotic dynamics and increase or decrease with the variation of nonlinear parameter of the system, such as those with similar injection angles~\cite{Custodio2012}. This is what has been found in soft Lorenz gases, a type of billiard-like system applicable for the description of transport properties in two-dimensional electronic materials~\cite{Gil-Gallegos2019}, where the breathing of chaos manifests as tongue-like' structures across the parameter space
~\cite{Klages2019,Toivonen2025,Toivonen2025arXiv}. There, the tongue-like structures correspond to the existence of different types of unstable periodic orbits, or quasi-ballistic trajectories in configurational space related to diffusive dynamics.

\begin{figure}
    \centering
    \includegraphics[width=0.9\columnwidth]{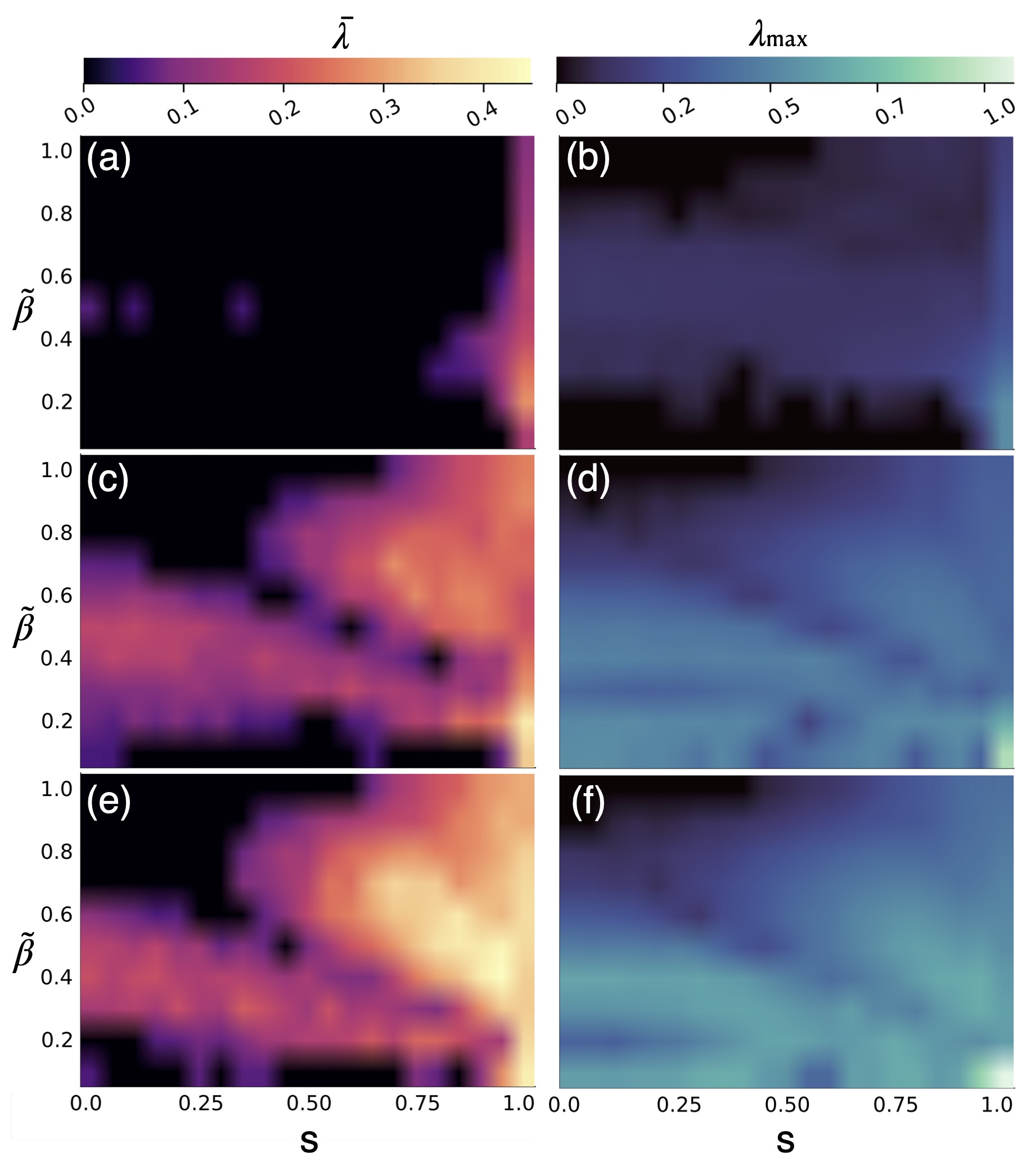}
    \caption{MLE for a random set of initial conditions as in Fig.~\ref{fig:4} with $E=0$, but as a function of $s$ and $\tilde{\beta}$. (Left) $\bar{\lambda}$ and (right) $\lambda_{\text{max}}$ with $h$ (from top to bottom): (a)-(b) $h=2$, (c)-(d) $h=4$, and (e)-(f) $h=6$.}
    \label{fig:6}
\end{figure}

Next, we explore the onset of chaos when we fix the geometry of the squircle and change the hardness. In Fig.~\ref{fig:6}, we show the free semi-axis $\tilde{\beta}$ vs. squareness $s$ space and study how it changes with several hardness values. This space notably demonstrates the dependence of the billiard dynamics on the variation of exclusively geometric parameters. Chaos becomes predominant on average as the boundary approaches the rectangular limit, in agreement what we have already observed in the $\tilde{\beta}-h$ space. In this scenario, where the roundness of corners is not changing but just the elongation of the billiard, the regularity-chaos-regularity transitions are not as pronounced though. We observe, however, `bubbles' of regularity where both the average and maximum MLE decrease close to zero. They correspond to the regular bands in Fig.~\ref{fig:5} that separate the circular from the squareness behavior, but are seen from a different cut of the parameter space. Similarly, in the right column of Fig.~\ref{fig:6}, essentially the same information is recovered, except that higher Lyapunov exponents are expected as hardness increases. This information is consistent with the previous parameter space in Fig.~\ref{fig:5}.

\begin{figure}
    \centering
    \includegraphics[width=0.9\columnwidth]{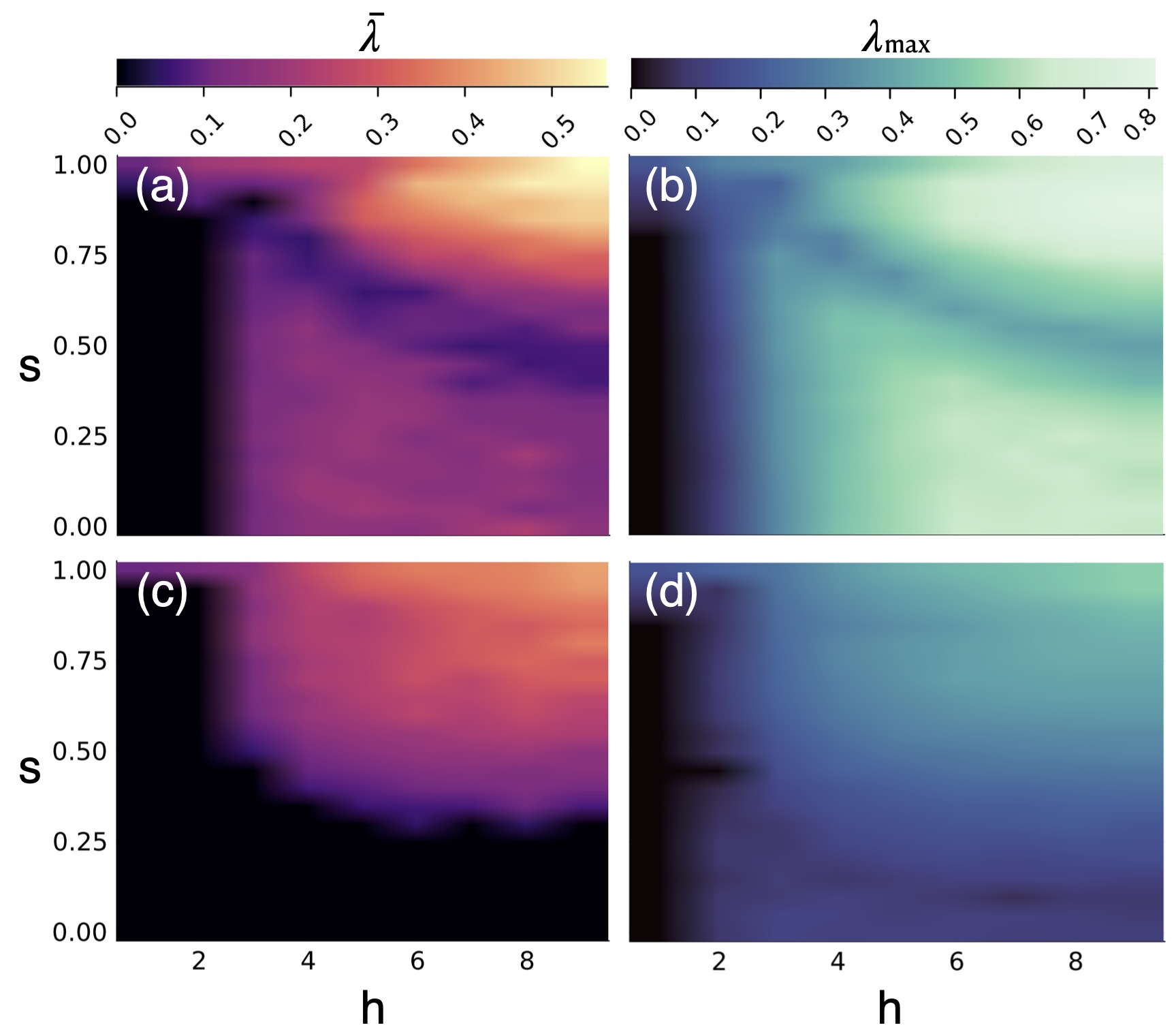}
    \caption{MLE for a random set of initial conditions as in Fig.~\ref{fig:5}, but as a function of $s$ and $h$, (left) $\bar{\lambda}$ and (right) $\lambda_{\text{max}}$. The values of the free semiaxis are fixed as: (a)-(b) $\tilde{\beta}=0.4$, (c)-(d) $\tilde{\beta}=0.8$ with $E=0$ for both cases.}
    \label{fig:8}
\end{figure}

Finally, in Fig.~\ref{fig:8}, the parameter space for $h$ and $s$ is shown, considering only two values of the free semi-axis $\tilde{\beta}$. In Fig.~\ref{fig:8} (a), it is observed how the increase in hardness, although less noticeable also generates regularity-chaos-regularity transitions, as well as the separation between the circular and square chaotic regions, which is consistent with what has been discussed so far. According to Fig.~\ref{fig:8} (b), these transitions are no longer present for a higher value of $\beta$. This is noteworthy because it provides information about the stabilizing effect of the geometric parameter $\beta$ (either stretching or reducing the available phase space) in relation to the boundary's squareness and the billiard’s hardness. The above is even more evident when considering the maximum MLE in Fig.~\ref{fig:8}. As shown in Fig.~\ref{fig:8}  (b), there are transition regions where the maximum value of the MLE can significantly vary , ranging from $\lambda_{\text{max}} = 0$ to  $\lambda_{\text{max}} \approx 0.8$. According to Fig.~\ref{fig:8} (d), this no longer occurs if the value of $\tilde{\beta}$ increases, as the parameter space is more uniform. Consequently, less chaotic dynamics are anticipated for low hardness and $s$ values, whereas the opposite holds for high $h$ and $s$ values.

A notable aspect of visualizing the previous parameter spaces is that they serve as a guide when constructing a billiard. Suppose one intends to study particle dynamics so that regularity prevails (or conversely). In that case, the combinations of geometry and hardness parameters can be chosen so that most initial conditions evolve following the expectations. This comprehension of a system’s dynamics holds significant relevance in chaos control research~\cite{Dube2000}.

\section{Conclusions}
\label{sec:6}

We have studied the dynamics of an anisotropic soft squircle billiard by solving its Hamilton equations and considering a two-dimensional potential well modeled by an error function. We have identified the emergence of chaos and regularity across the system's parameter space, constituted by the free semiaxis and squareness that control the geometry and the hardness, by computing the Poincar\'e surface of sections and the maximal Lyapunov exponent for different combinations of parameters for a given set of random initial conditions of the same energy. Although it is known that hardness introduces chaos in the billiard, even for those that are regular in the hard-wall limit, we have discovered that the relationship between geometric parameters and smoothness is  nonlinear in the case of squircle geometry. 

We have found that the circular and squareness features of the billiard dominate the emergence of chaos in different regions of the parameter space, such that in the intermediate squircle they converge. However, a regularity band maintains the separation between them.  Deforming the geometry of the billiard has the effect of stabilizing the dynamics, so this band manifests as regular `bubbles' in the parameter space. Likewise, the emergence of stability islands in the geometric parameter space and their nonlinear breakdown were observed as functions of the hardness parameter. Specifically, the variation of hardness introduces chaotic stripes in the parameter space, producing an intermittence between chaos and regularity, or breathing of chaos. This behavior is still present as a function of geometrical parameters. We believe this is produced by the rearrange of phase space structures associated with sets of orbits such as in soft Lorenz gases. The interplay between the squareness, which controls the roundness of the corners, and the billiard’s hardness, enriches the phenomenon. However, an the ultimate dynamical explanation is still undisclosed, and several questions remains, as how the phenomenology depends on the shape of the potential.

Another possibility would be the emergence of `shrimp'-like structures originally identified in the Hénon map~\cite{Gallas1993,Gallas1994}. Shrimp-like structures are a generic feature of a broad class of dissipative systems with chaotic attractors, where a window of stable periodicity of a given period, surrounded by a chaotic sea, looks like a `shrimp' within a two-dimensional parameter diagram~\cite{Oliveira2011,Stoop2012,Lim2024,Pati2024}. It has been looked for shrimp-like structures in conservative systems. However, continuous time evolution represents an issue for attaining the necessary resolution for observing them~\cite{Nascimento2011}. Although we do not attain numerical resolution enough to evince a connection between the stripes and shrimp-like structures, we believe soft billiard systems present an interesting opportunity for further research of the latter.

Through its geometric properties, the squircle billiard introduces additional freedom in the parameter space. This enables the study of the persistence of the chaos-regularity-chaos dynamics in terms of geometric variations, posing a novel model as a valuable resource for investigating confined colliding particles with irregular boundaries or the effect of rounded corners in a rectangular billiard. Our results might help characterize experimental systems such as microwave resonators, or to provide a guidelines for the experimental control of phase-space. Furthermore, extending this model to the quantum domains is a natural subsequent step that we plan to address in the near future.  

\section*{Acknowledgements}
AGA and MABM acknowledge financial support from CONAHCYT/SECIHTI No. CBF2023-2024-1765. MABM also acknowledges the financial support from the Marcos Moshinsky Fellowship Program and the PIPAIR 2024 project from the DAI-UAM. We thank J. G. Hirsch and the support of the Computation Center—ICN-UNAM, particularly E. Palacios, L. Díaz, and E. Murrieta. We also thank E. Ben\'itez Rodr\'guez for providing valuable comments on this paper, and to R. Klages for his insight on soft Lorenz gases.

\newpage

\bibliography{SB.bib}

\end{document}